\documentclass{eas}
\usepackage{graphicx}
\begin{document}

%%-----------------------------
%%      the top matter
%%-----------------------------
\title{Conditions for star formation in nearby AGN and QSO hosts observed with near-infrared integral-field spectroscopy} 
\runningtitle{IFS studies of star formation in AGNs and QSOs}
\author{Gerold Busch}\address{I. Physikalisches Institut der Universit\"at zu K\"oln, Z\"ulpicher Str. 77, 50937 K\"oln, Germany}
\author{Nastaran Fazeli}\sameaddress{1}
\author{Semir Smajic}\sameaddress{1}\secondaddress{Max-Planck-Institut f\"ur Radioastronomie, Auf dem H\"ugel 69, 53121 Bonn, Germany}
\author{Andreas Eckart}\sameaddress{1,2}
\author{Lydia Moser}\sameaddress{2,1}
\author{M\'onica Valencia-S.}\sameaddress{1}
\begin{abstract}
Integral-field spectroscopy in the near-infrared (NIR) is a powerful tool to analyze the gaseous and stellar distributions and kinematics, as well as the excitation mechanisms in the centers of galaxies. The unique combination of NIR and sub-mm data at comparable high angular resolution, which has just been possible with SINFONI and ALMA, allows to trace warm and cold gas reservoirs. Only the NIR gives an unobscured view to the center and allows to study the conditions and impact of star formation in the centers of galaxies in a spatially resolved way. Here, we present recent studies of nearby Seyferts and low-luminosity QSOs performed by our group.
\end{abstract}
\maketitle
%%-----------------------------
%%      your text
%%-----------------------------
\section{Introduction}

Near-infrared (NIR) integral-field spectroscopy (IFS) is a powerful tool to study the star formation and nuclear activity of dust-enshrouded galactic nuclei. Integral-field spectroscopy provides spectra for every spatial pixel which allows the extraction of gaseous and stellar distribution and kinematics, as well as the study of excitation mechanisms in a spatially resolved way. The NIR is less affected by dust absorption than the optical and mainly traces the mass-dominating old stellar population. An example is our study of NGC 7172, formerly classified as Seyfert-2 galaxy, where we detect broad hydrogen emission lines (indicative for a broad-line region) behind a dust-lane which hides the real center in the optical light (Smaji\'c \etal\ \cite{smajic12}).

We present our recent studies that are based on observations with the NIR IFS SINFONI (Eisenhauer \etal\ \cite{eisenhauer03}, Bonnet \etal\ \cite{bonnet04}) which has a typical spectral resolution of $\approx 100\,\mathrm{km/s}$ and a spatial resolution of $0.8\,$arcsec (seeing-limited) or $100\,$mas (adaptive optics) resp.

First, we briefly discuss the importance of NIR observations for the interpretation of sub-mm ALMA data. Then, we show first results from a NIR IFS study of the nuclear star-formation properties of the Seyfert-2 galaxy NGC 7496. Finally, we discuss the impact of circumnuclear star formation in low-luminosity QSOs on the black hole mass ($M_\mathrm{BH}$) - bulge luminosity ($L_\mathrm{bulge}$) relation.

\section{Nearby Seyferts observed with SINFONI and ALMA}

The NUclei of GAlaxies (NUGA) project (Garc\'ia-Burillo \etal\ \cite{garciab03}) comprises a sample of two dozens of nearby galaxies which span the whole range of nuclear activities.
Two galaxies from the NUGA sample have recently been observed in the NIR with SINFONI and in the sub-mm with ALMA. 

In NGC 1433, we can determine the real position of the AGN only from the NIR data. A steep velocity gradient in the center (in ALMA/CO and SINFONI/warm H$_2$) is observed that could be interpreted as molecular outflow. However, from NIR emission line diagnostics, we conclude that the nuclear star formation rate is low compared to other low-luminosity AGN. This means that the outflow would be essentially triggered by the AGN alone (Combes \etal\ \cite{combes13}). An alternative explanation is a nuclear disk (Smaji\'c \etal\ \cite{smajic14}).

In NGC 1566, we find a star-formation region in $\sim 2\,\mathrm{arcsec}$ distance from the nucleus. From NIR emission lines, we estimate the star formation rate in the circumnuclear disk to be $8\times 10^{-3}\,M_\odot\,\mathrm{yr}^{-1}$.
The molecular gas content from NIR agrees with the gas mass from CO ($10^7 - 10^8\,M_\odot$; Combes \etal\ \cite{combes14}). We conclude that the nuclear region has a large reservoir of molecular gas that is not used efficiently. NIR line ratios show that the gas infall is accompanied by young star formation ($<9\,\mathrm{Myr}$; Smaji\'c \etal\ \cite{smajic15}).

\section{Circumnuclear star formation in NGC 7496}

The central $\sim 300\,\mathrm{pc}$ of NGC 7496 were observed with ESO-SINFONI in $H+K$-band. The galaxy is a nearby barred spiral (SBbc), at a distance of $19.4\,\mathrm{Mpc}$ ($z=0.0055$). 

\begin{figure}
\includegraphics[width=\linewidth]{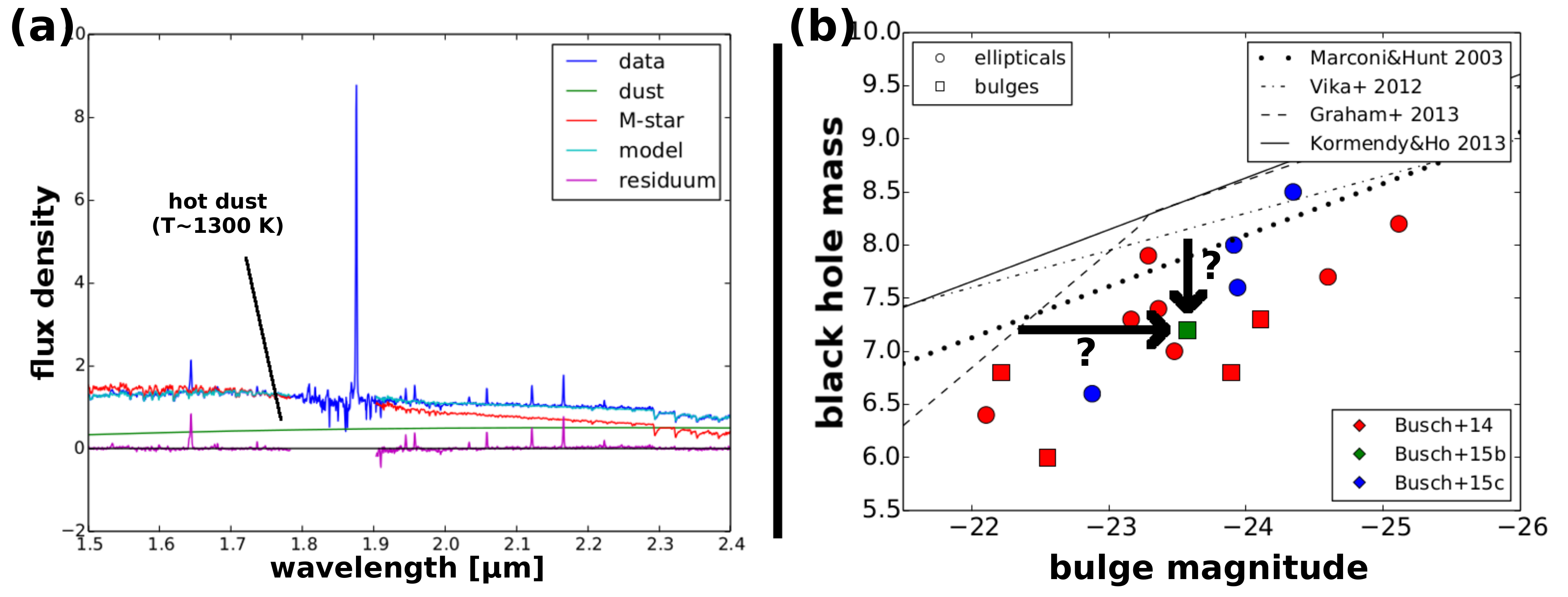}
\caption{\emph{Left}: continuum fit of the Seyfert galaxy NGC 7496. The hot dust component with $T\sim 1300\,\mathrm{K}$ is apparent. \emph{Right}: (Black hole mass) - (bulge luminosity) relation. LLQSOs do not follow these relations.}
\label{figure}
\end{figure}

Previous studies of this AGN in the mid-infrared show a very hot dust temperature ($T \sim 1600\,\mathrm{K}$; Prieto \etal\ \cite{prieto02}). In the NIR, the continuum emission fit also shows a dust temperature of around $T \sim 1300\,\mathrm{K}$, which is high for a Seyfert-2 (Fig.~\ref{figure} (a)). We see indications that the nucleus might power a very faint and narrow broad line region ($\mathrm{FWHM}_\mathrm{broad}<1500\,\mathrm{km}\,\mathrm{s}^{-1}$; Valencia-S. \etal\ \cite{valencias14}). 

This motivates us to study the gas content and dynamics in the center of this galaxy. The spectrum shows a variety of strong emission lines. In a forthcoming paper (Fazeli \etal), we investigate the properties of these emission lines and their ratios to trace circumnuclear star formation regions and reveal their excitation conditions, morphology, and stellar content.

\section{Star formation and $M_\mathrm{BH}-L_\mathrm{bulge}$ relations in low-luminosity QSOs}

To fill the gap between nearby Seyfert galaxies and high-$z$ quasar samples, we also study low-luminosity QSOs (LLQSO). The LLQSO sample (e.g., Bertram \etal\ \cite{bertram07}, Busch \etal\ \cite{busch13}) contains 99 nearby type-1 quasars with redshift $0.02\leq z \leq 0.06$ and connects nearby AGN and more distant quasars in many properties (Moser \etal\ \cite{moser13}). 
Owing to their larger distance, the angular scale is less by a factor of $\sim 10$ (which is still resolvable in detail). However, their bolometric luminosities and accretion rates are higher by several magnitudes which makes them interesting to study (see also new survey: www.cars-survey.org).

Using black hole masses from single-epoch spectroscopy and bulge luminosities from bulge-disk-bar-AGN decomposition (using BUDDA, Gadotti \cite{gadotti08}) of deep NIR images (from SOFI@NTT and LUCI@LBT), we show that LLQSOs do not follow $M_\mathrm{BH}-L_\mathrm{bulge}$ relations established for inactive galaxies (see Fig.~\ref{figure} (b); Busch \etal\ \cite{busch14}, Busch \etal\ \cite{busch15c}). Possible explanations are (1) bulges which are over-luminous due to star formation/young stellar populations or (2) undermassive black holes that are in a growing phase. Understanding this offset has important implications on the co-evolution of BHs and bulges. Sun \etal\ (\cite{sun15}) investigate the ``flow patterns'' of BHs and their host galaxies in the $M_\mathrm{BH}-M_*$ relation and show that galaxies that are off-set from the relation have evolutionary vectors anti-correlated with their position, indicating that they are moving back towards the relation.

To understand the behaviour of LLQSOs in these relations, detailed studies of nuclei of LLQSOs are performed. In a pilot study, we observed HE 1029-1831 with SINFONI, seeing-limited and with adaptive optics (Busch \etal\ \cite{busch15a}, Busch \etal\ \cite{busch15b}, Moser \etal\ \cite{moser15}). We detect a star-forming ring around the nucleus with radius $\sim 240\,\mathrm{pc}$. Comparing observables like the equivalent width of Br$\gamma$ or the supernova rate (derived from [Fe\textsc{ii}] flux) to models from \textsc{Starburst99}, we conclude that the ring experienced a starburst with time scale $\tau \sim 50-100\,\mathrm{Myr}$ that began $\sim 100-200\,\mathrm{Myr}$ ago. The bulge luminosity is dominated by this intermediate-age stellar population which explains the offset in the $M_\mathrm{BH}-L_\mathrm{bulge}$ relation, at least in this particular case.

\section{Conclusions and outlook}

We discussed the power of near-infrared integral-field spectroscopy to analyze nuclear activity and star formation in nearby AGN and low-luminosity QSOs. Furthermore, we showed that these observations are crucial for a thorough analysis of sub-mm ALMA observations of the gas content. 

A bigger sample of nearby Seyferts will allow us to probe possible relations between nuclear activity and nuclear star formation in great detail. More observations of LLQSOs are necessary to robustly determine the impact of nuclear star formation on the position of these objects in $M_\mathrm{BH}-L_\mathrm{bulge}$ relations.

\paragraph*{Acknowledgements}
This work is carried out within the Collaborative Research Centre 956, sub-project A2, funded by the DFG. 
G.~Busch and N.~Fazeli are supported by Bonn-Cologne Graduate School of Physics and Astronomy.
%%-----------------------------
%%      your bibliography
%%-----------------------------

\end{document}